(57) **ABSTRACT**

Due to an ever growing shortage of conventional energy sources, there is an increasingly intense interest in harnessing solar energy. The instant invention can contribute to the goal of achieving environmentally clean solar energy to be competitive with conventional energy sources. A novel method is described for manufacturing a transparent sheet with an embedded array of mirrored spheroidal micro-balls for use in a solar energy concentrator, and analogous applications such as optical switches and solar rocket assist. The micro-balls are covered with a thin spherical shell of lubricating liquid so that they are free to rotate in an almost frictionless encapsulation in the sheet. Novel method and apparatus are presented for producing the preferred embodiment of a close-packed monolayer of the array of mirrored micro-balls.

**22 Claims, 2 Drawing Sheets**

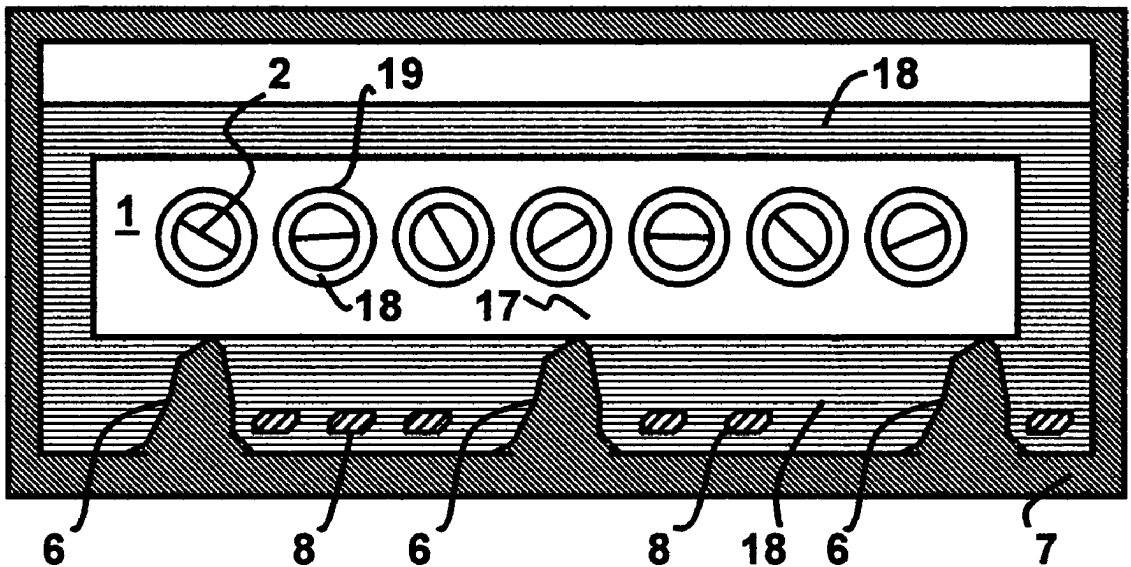

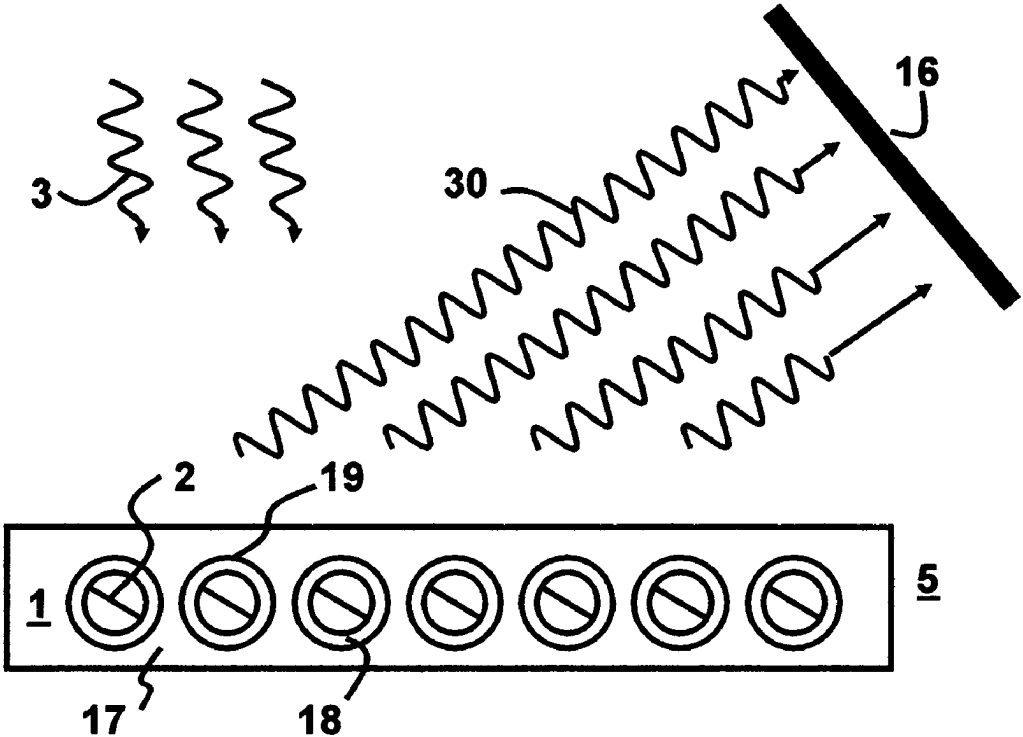

Fig. 1

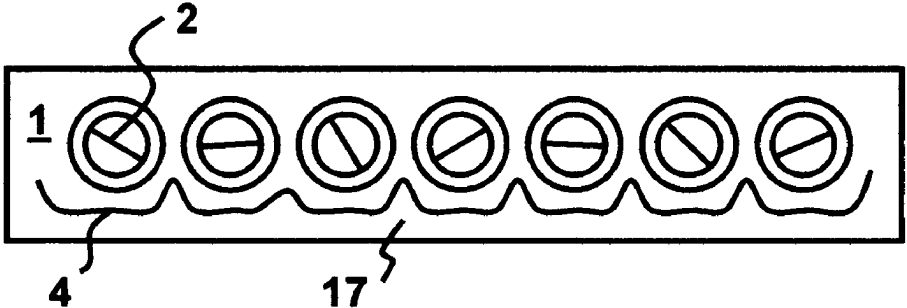

Fig. 2

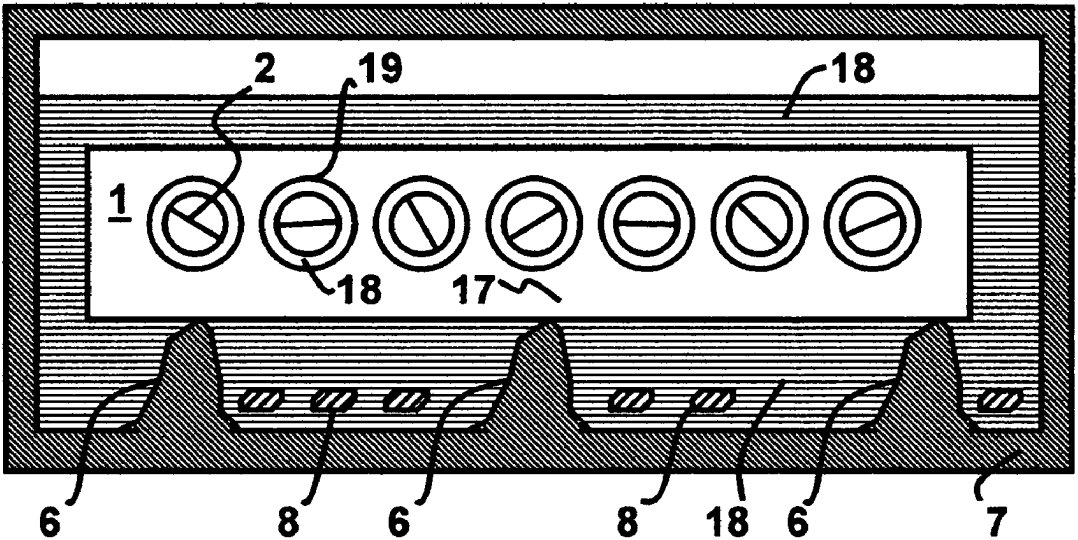

Fig. 3

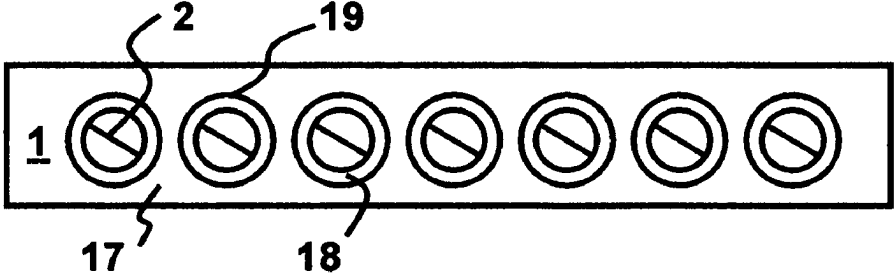

Fig. 4

# MANUFACTURE OF AND APPARATUS FOR NEARLY FRICTIONLESS OPERATION OF A ROTATABLE ARRAY OF MICRO-MIRRORS IN A SOLAR CONCENTRATOR SHEET

## INCORPORATION BY REFERENCE

The following owned in common U.S. patents, allowed patent applications, and pending patent applications are fully incorporated herein by reference:

U.S. Pat. No. 6,612,705, by Mark Davidson and Mario Rabinowitz, "Mini-Optics Solar Energy Concentrator" issued on Sept. 2, 2003.

U.S. Pat. No. 6,698,693 by Mark Davidson and Mario Rabinowitz, "Solar Propulsion Assist" to issue on Mar. 2, 2004.

U.S. Publication #2003-0202235-A1, by Mario Rabinowitz and Mark Davidson, "Dynamic Multi-Wavelength Switching Ensemble" allowed on Oct. 6, 2003.

U.S. Publication #Not Designated Yet, by Mario Rabinowitz, "Spinning Concentrator Enhanced Solar Energy Alternating Current Producton" is Pending.

## BACKGROUND OF THE INVENTION

This invention provides a low cost means for achieving affordable solar energy by greatly reducing the cost of solar concentrators which increase (concentrate) the density of solar energy incident on the solar energy converter. A limiting factor in the utilization of solar energy is the high cost of energy converters such as photovoltaic cells. For example, for the purpose of generating electricity, a large area of expensive solar cells may be replaced by a small area of high-grade photovoltaic solar cells operating in conjunction with the inexpensive intelligent micro-optics of this invention. Thus the instant invention can contribute to the goal of achieving environmentally clean energy on a large enough scale to be competitive with conventional energy sources.

The 1979 Gyricon U.S. Pat. No. 4,143,103 of Sheridon, entitled "Method of Making a Twisting Ball Display" and the 2002 U.S. Pat. No. 6,441,946 of Sheridon, entitled "Swollen Gyricon Displays and Method of Making Same" are exclusively concerned with Displays. There appears to be no mention of any other application than Displays, either specifically or by general statement. In these Sheridon patents, no mention is made of a mirror in the gyricon balls, nor is there any mention of specular reflection as would be obtained from a mirror. On the contrary, diffuse reflection needs to be increased from the balls so the Gyricon display may easily be observed from all angles. Certainly there is no anticipation of a solar concentrator application, solar propulsion assist, optical switching or any other micro-mirror application. Furthermore, a uniform monolayer(s) of micro-mirrored balls are preferred in these applications, whereas the Sheridon patents only teach a random dispersion of non-mirrored Gyricon balls.

The instant invention is primarily concerned with the manufacture of sheets that hold solar concentrator micro-mirrors for nearly frictionless rotation. However, it has broader applications wherever mirrors are used for micro-mirror focusing applications such as for solar propulsion assist and optical switching.

Definitions

"Binder" refers herein to a material additive that is used to promote solidification, provide mechanical strength, or to ensure uniform consistency.

"Concentrator" as used herein in general is a micro-mirror system for focusing and reflecting light. In a solar energy context, it is that part of a Solar Collector system that directs and concentrates solar radiation onto a solar Receiver.

"Dielectric" refers to an insulating material in which an electric field can be sustained with a minimum power dissipation.

"Elastomer" is a material such as synthetic rubber or plastic, which at ordinary temperatures can be stretched substantially under low stress, and upon immediate release of the stress, will return with force to approximately its original length. Silicone elastomers have exceptional ability to withstand ultraviolet light degradation.

"Electret" refers to a solid dielectric possessing persistent electric polarization, by virtue of a long time constant for decay of charge separation.

"Focusing planar mirror" is a thin almost planar mirror constructed with stepped varying angles so as to have the optical properties of a much thicker concave (or convex) mirror. It can heuristically be thought of somewhat as the projection of thin equi-angular segments of small portions of a thick mirror upon a planar surface. It is a focusing planar reflecting surface much like a planar Fresnel lens is a focusing transmitting surface. If a shiny metal coating is placed on a Fresnel lens it can act as a Fresnel reflector.

"Packing fraction" herein refers to the fraction of an available area occupied by the collection (ensemble) of rotatable elements.

"Plasticizer" as used herein refers specifically to a dielectric plasticizer fluid that is absorbed by an elastomer thereby causing it to swell thus creating a spherical shell around each of the micro-mirror balls which do not substantially absorb the plasticizer. More commonly plastizers are added to a material to make it softer, more flexible, or more moldable.

"Receiver" as used herein in general is a system for receiving reflected light. In a solar energy context, it receives concentrated solar radiation from the micro-mirror assembly for the conversion of solar energy into more conveniently usable energy such as electricity.

"Silicone" as used herein refers to a heat-stable, rubber-like elastomer that is a water repellent, semiorganic polymer of organic radicals attached to silicon containing molecules, such as dimethyl silicone. Silicone elastomers are an excellent material within which to embed the mirrored balls or cylinders, because of their durability with respect to ultraviolet light, among other reasons.

"Thermoplastic" refers to materials with a molecular structure that will soften when heated and harden when cooled. This includes materials such as vinyls, nylons, elastomers, fuorocarbons, polyethylenes, styrene, acrylics, cellulosics, etc.

"Zeolyte" is a group of hydrous tectosilicate minerals such as analcime, chabazite, natrolite, and stilbite usually characterized by an aluminosilicate tetrahedral framework and ion-exchangeable large cations permitting reversible dehydration. Zeolytes are an excellent material to add to the plastizer liquid to help keep it clean and if necessary deionized.

## SUMMARY OF THE INVENTION

It is an object of the present invention to develop a method of making a solar concentrator sheet in which the micro-mirrors may rotate in an almost frictionless encapsulation in a sheet.

An aspect of this invention is to provide apparatus for nearly frictionless rotation of micro-mirrored balls i.e. spheres or cylinders.

It is a further aspect of the present invention to achieve a solar concentrator that can tolerate a range of micro-mirrored ball sizes.

Another aspect of this invention is to control the size of the cavity which encapsulates the micro-mirrored balls.

Another aspect of this invention is to make the bottom of the sheet more accessible to the plasticiser so that swelling may take place more uniformly.

Another object of the present invention is to enable the encapsulation of an approximately uniform monolayer of micro-mirrored balls.

These and other objects of the instant invention are achieved herein by swelling the solar concentrator sheet with one or more liquids. A single liquid may function both to swell the concentrator sheet and lubricate the micro-mirrored balls. As will be described, a combination of two or more liquids may distribute the functions of swelling and lubrication more precisely.

There are many aspects of this invention for the method of manufacture of sheets that hold mirrored balls to be used in solar concentrators and analogous applications. All the aspects apply to the same objective of being technically sound, economically viable, practical, and efficient in this goal.

Other objects and advantages of the invention will be apparent in a description of specific embodiments thereof, given by way of example only, to enable one skilled in the art to readily practice the invention singly or in combination as described hereinafter with reference to the accompanying drawings. In the detailed drawings, like reference numerals indicate like components.

## BRIEF DESCRIPTION OF THE DRAWINGS

FIG. 1 is a cross-sectional view of a micro-optics solar concentrator with the micro-mirrors focusing the incident sunlight onto a receiver.

FIG. 2 is a cross-sectional view showing micro-mirrored balls on a tray in an elastomer sheet that is to be cured.

FIG. 3 is a cross-sectional view of an oil bath showing in cross-section an immersed swollen, cured elastomer sheet on pillars, with spherical shells of lubricating liquid surrounding each ball.

FIG. 4 is a cross-sectional view of a micro-optics sheet showing the mirrored balls covered with a lubricating liquid film in enlarged cavities.

## DETAILED DESCRIPTION OF THE PRESENTLY PREFERRED EMBODIMENTS

There are many tradeoffs in manufacturing the transparent sheet that holds the micro-mirrored balls and cylinders for solar energy and analogous applications. One tradeoff is between materials costs versus manufacturing costs. Another tradeoff is between small and large volume fabrication. Another tradeoff is between materials costs and power requirements. The conclusions reached today may well have to be changed in the light of new costs and new developments in the future. As is described here in detail, the apparatus and manufacturing objectives of the instant invention may be accomplished by any of a number of ways separately or in combination, as taught herein.

FIG. 1 is a cross-sectional view of a micro-optics ensemble 5 of an individually rotatable monolayer of elements 1 (balls) showing the embedded micro-mirrors 2 which focus the incident sunlight 3 as concentrated light of the reflected wave 30 onto a receiver 16. The receiver 16 as used herein denotes any device for the conversion of solar energy such as electricity, heat, pressure, concentrated light, etc. One monolayer of elements 1 as shown is a preferable embodiment, though more than one layer may be used. For some purposes a random dispersion of elements 1 is acceptable. Refer to U.S. Pat. No. 6,612,705 of which one inventor of this instant invention is the co-inventor, for some of the ways in which the micro-mirrors 2 can be aligned to focus incident light. The elements 1 may be spherical or cylindrical, and for convenience will sometimes be referred to as balls. Spherical elements are preferable for two-axis tracking but have a smaller packing fraction than cylindrical elements whose packing fraction can approach 1. The rotatable elements 1 have at least one transparent hemisphere (or hemicylinder as the case may be) and are dispersed in an elastomer sheet 17.

The elements 1 with embedded micro-mirrors 2 are surrounded by and suspended in a dielectric lubricating liquid 18 that allows them to rotate freely inside an enlarged cavity 19 that encapsulates them. It is preferable to utilize a liquid 18 whose index of refraction matches the clear hemisphere or clear hemicylinder, and it should have the same density as element 1 to minimize buoyant forces. The index of refraction of the sheet 17, the liquid 18, and the optically transmissive upper portion of elements 1 should all be approximately equal. The elements 1 should be roughly balanced to minimize gross gravitational orientation.

The minimum diameter of elements 1 can be assessed from the Rayleigh limit

$$d = \frac{0.61\lambda}{n \sin u} \sim 10\lambda,$$

where d is the minimum diameter of elements 1, $\lambda \sim 4000$ Å is the minimum visible wavelength, n is the index of refraction $\sim 1$ of element 1 (the medium in which the incident light is reflected), and u is the half angle admitted by elements 1. Thus d $\sim 40{,}000$ Å ($4 \times 10^{-6}$ m, i.e. 4 microns) is the minimum diameter of elements 1. A maximum for the diameter of elements 1 is $\sim 10$ mm. A preferable range is 50 microns to 400 microns.

If the focusing planar mini-mirrors concentrate the solar radiation by a factor of 100, the total increase in power density reaching the collector would be 100 times greater than the incident power of the sun. Thus the receiver area need be only $\sim 1\%$ the size of one receiving solar radiation directly. Although the total capital and installation cost of this improved system may be more than 1% of a direct system, there will nevertheless be substantial savings.

FIG. 2 is a cross-sectional view showing elements 1 with equatorial micro-mirrors 2, which elements 1 are on a tray 4 in an elastomer sheet 17 that is to be cured. The uncured elastomer 17 is molten and is preferably transparent silicone rubber. The tray 4 holds the elements 1 in place during curing, and is preferably dissolved in the process. In the case when the tray 4 is dissolved away during curing, then it may completely surround the elements 1 i.e. it may also be on the top of elements 1, as well as a bottom support as shown. The tray 4 can have a variety of configurations ranging from a large flat box, to an egg carton-like arrangement, to little individual cubicles each holding a ball. Vibration may be employed to get one ball per cubicle prior to introduction of the ball-filled tray into the elastomer mold. If a large flat box is employed as the tray 4, the elements 1 may be pre-coated with the uncured elastomer to maintain a minimum separation between them. When the tray 4 is not dissolved, it may be perforated or dismantled so that it may be eliminated from proximity to the elements 1 as the elastomer 17 thickens. The tray 4 may then remain in place away from the elements 1, or completely removed as desired.

For optical reasons, it is preferable to have the elements 1 as close as possible to the top surface through which light is incident and returns. For necessary mechanical strength of the sheet, the bottom of the elastomer 17 may be far from the balls. So the elastomer 17 is preferably thin at the top, and may be thick at the bottom below the balls to give it mechanical strength. The elastomer 17 may be made in modular segments or if made larger than the desired area, after the elastomer 17 solidifies, it is cut into sheets. At this stage the elements 1 are being held rigidly captive in the elastomer 17.

A single monolayer of elements 1 is preferred as this reduces manufacturing costs, and light need only traverse a thin layer of material in reaching and reflecting from the micro-mirrors 2. If desired, more than one monolayer of elements 1 may be similarly formed by the concurrent use of two or more trays 4, one below the other. If this is done, it is preferable to have successive layers jogged to increase the packing fraction seen by the incident light. Alternatively, the elastomer sheet 17 can be formed by thoroughly mixing the elements 1 with the uncured, liquid, optically transparent material and then curing. The optically transparent material is then cured by heating, and/or chemically. An example of an uncured elastomer that can be used is Dow Corning Sylgard 182 which is cured by rapid heating from ambient temperature to about 140 Centigrade degrees for a duration of about 10 to 15 minutes.

FIG. 3 is a cross-sectional view of a dielectric plasticizer lubricating liquid 18 in a closed container 7 showing in cross-section an immersed cured elastomer sheet 17, which has been swollen by the infiltration of liquid 18. Thus spherical shells of lubricating liquid 18 surround each element 1 inside an enlarged cavity 19. Zeolytes 8 are in the liquid 18 to help keep it clean and if necessary deionized. Pillars 6 support the sheet 17 to make the bottom of the sheet 17 more accessible to the plasticiser liquid 18 so that swelling may take place more uniformly. This is overlooked in the Gyricon sheet Sheridon U.S. Pat. Nos. 4,143,103 and 6,441,946, and the liquid 18 has little access to the bottom of the sheet 17. The Gyricon sheet is shown laying at the bottom of the container with plasticiser liquid shown only on top of it, limiting the ingress of liquid.

The immersed elastomer sheet 17 has soaked up the dielectric plasticizer lubricating liquid 18, causing it to expand, creating spherical or cylindrical cavities (as the case may be) around each element 1. The liquid 18 may be a single molecular species which can infiltrate the sheet 17 or a combination of different molecular species which in concert can infiltrate the sheet 17. At ambient temperature, time periods of between 5 to 10 hours provide sufficient expansion to form an adequate spherical or cylindrical shell of liquid 18 around each element 1. This enables the elements 1 to float and rotate freely, Now that the elements 1 (balls) have been liberated to float and rotate freely, they can respond to coupling fields for alignment and focusing as described in U.S. Pat. No. 6,612,705 of which one inventor of this instant invention is the co-inventor. For use with the elastomer Sylgard 182, the dielectric plasticizer liquid 18 can be silicone oil, such as Dow Corning 10 Centistoke 200 oil. Dow Corning 10 Centistoke 200 oil may also be used as the plasticizer liquid 18 with Stauffer and Waker's elastomer V-53.

Not all candidate liquids 18 have the entire combination of desirable properties of not causing too much swelling, being a good dielectric, being a good lubricant with low viscosity, being transparent, being non-toxic, etc. When this occurs, it is preferable to use a combination of two or more liquids. If the one that primarily controls the swelling (but does not have the other desirable properties) is more volatile, then it can be displaced by the other liquid(s) as it is removed by evaporative heating. If necessary, after the desired degree of swelling has been achieved, the swollen sheet can be heated in a bath containing only the liquid(s) 18 that have the desired properties.

There are a number of liquids that may be used in combination. Partially fluorinated materials may be used such as 3M HFE 7100, which is a partially fluorinated hydrocarbon made by Minnesota Mining and Manufacturing. Two others are Isopar L or Isopar M, which are aliphatic hydrocarbons made by Ashland Chemicals. Another good swelling liquid is Freon TF, a partially fluorinated polyethylene. The rate and degree of swelling is a function of both the properties of the liquid 18 and the elastomer sheet 17. Usually a lower elastic modulus sheet 18 swells at a higher rate than one with a higher elastic modulus i.e. one that requires higher mechanical force to expand or compress it.

Let us consider the ratio of the volume of material of sheet 18 (e.g. elastomer) to the volume of cavities 19 containing elements 1. For the sake of brevity and simplicity, let us consider the elements 1 to be spherical balls. The minimum possible ratio of the volume of the sheet to the volume of the balls is the ratio of the volume of a cube of side length 2r to the volume of the inscribed sphere of radius r. This is:

$$\frac{V_{sheet}}{V_{balls}} = \frac{(2r)^3}{\left(\frac{4}{3}\pi r^3\right)} = 1.91 \approx 2.$$

Therefore when mixing the material of the sheet 18 with the balls 1, the ratio of the two volumes must be at least 1.91 or about 2 or more.

A lubricating layer of liquid 18 that is about a 10% radial liquid coating changes the ratio of the volume of material of sheet 18 to the volume of spherical balls to:

$$\frac{V_{sheet}}{V_{balls}} = \frac{(2[1.1r])^3}{\left(\frac{4}{3}\pi r^3\right)} 2.54.$$

FIG. 4 is a cross-sectional view of a micro-optics ensemble 5 of an individually rotatable monolayer of elements 1 (balls) showing the embedded micro-mirrors 2 in a sheet 17, as the main constituent of a micro-optics solar concentrator. The elements 1 are covered with a lubricating liquid film 18 inside enlarged cavities 19 whose degree of swelling has been controlled.

Liquids that cause little or no swelling may be combined with the good swellers, for their desirable properties or to help regulate swelling. The relative percentages can easily be determined empirically. Preferably, the two sets of liquids are mutually soluble for uniformity of mixture, and so that they may be soaked up together. Such liquids include fully fluorinated materials such as perfluorooctane that causes little or no swelling of an elastomeric binder. Others are low molecular weight triglyceride liquids such as tributyrin and tricaproin. Even natural vegetable oils such as soybean oil, coconut oil, and walnut oil may be used, as long as care is exercised to keep them from becoming rancid. Walnut oil is an excellent, low viscostiy lubricant that can be obtained from crushed walnuts. The various grades of walnut oil can easily be gravity separated in vertical columns.

In order to have a high packing fraction of the micro-mirrors it is important to control the swelling of the sheet so that the balls remain as close together as feasible. The presently preferred degree of swelling of the cavity radius around each ball is between 2% and 15%. To minimize friction and concomitantly reduce the torque power requirement, it is imperative that at least a 1% radius coating of liquid lubricant completely coats each ball. If the sheet is removed from the dielectric plastizer lubricating liquid to stop the swelling process, prior to achievement of at least a 1% radial lubricant coating, it will be difficult to properly align the micro-mirrors. The degree of swelling can be controlled through use of a larger percentage non-swelling liquid mixed with a smaller percentage swelling liquid, and subsequent removal by evaporation of the swelling liquid.

It is preferable to make the elements **1** from a material that does not easily absorb the liquid plasticizer **18** or does so at a much lower rate than the cured elastomer sheet **17**. Glass is a preferable material for the elements **1**, both for this reason as well as for its resistance to ultraviolet degradation.

In addition to reducing rotational friction, another advantage of encapsulation of the elements **1** is that one can more easily tolerate a range of ball sizes. When a close range is desired, it is not necessary to force expensive manufacturing tolerances to make balls of all of about the same size. The balls can be sorted by sieves into size ranges.

Operational Issues, Advantages, and Alternatives

The optically transparent sheet in which the balls i.e. spheres or cylinders are imbedded, need not be an elastomer, though a silicone elastomer is presently preferred. The sheet may be made of plastics such as polyethylene, polystyrene or plexiglass. Encapsulation can be achieved with the encapsulant molten or dissolved in a volatile solvent. An uncured rigid material such as an epoxy can be used as the encapsulant binder provided that it is light transparent. It is imperative that the material of the sheet absorb the dielectric plastizer liquid lubricant more readily than do the balls in order to form the liquid filled cavities around each ball. When the material of the gyricon sheet binder is an elastomer, the spheres can be plastics such as polyethylene or polystyrene (neglecting ultraviolet degradation) which do not absorb the plasticizer as readily as elastomers.

When the material of the binder is a plastic, the balls must be of a material such as glass which does not absorb the liquid, or absorbs the dielectric liquid much less than does the plastic. Additional reasons for using glass balls are its superior durability with respect to ultraviolet light; its excellent optical properties such as high transparency; and its low cost. Though not as good as glass, silicone has high ultraviolet light durability as well as good electrical properties. Electrical cables impregnated with silicone additives better withstand electrical and water treeing. Other possible polymers that may prove promising are TPX (4-methyllpentene-1), Aurem (Polyimide), and SPS (Syndiotactic Polystyrene). An alternative approach to a single material for the sheet uses these and non-thermoplastic polymers in a laminated film form. PQ-100 (Polyquinoline) or Isaryl 25 can also be used in laminar form. Cross-linkable silicone or 1,4-polybutadiene based resin can be used to permeate the bottom of the sheet and fill unwanted voids.

While the instant invention has been described with reference to presently preferred and other embodiments, the descriptions are illustrative of the invention and are not to be construed as limiting the invention. Thus, various modifications and applications may occur to those skilled in the art without departing from the true spirit and scope of the invention as summarized by the appended claims together with their full range of equivalents.

The invention claimed is:

1. A method for fabricating lubricating receptacles containing encapsulated rotatable mirrored balls in an optically transmissive solar concentrator sheet by means of at least one infiltrating fluid, the method comprising the process of:
   a) distributing solid mirrored balls in a solidifiable mixture to form a sheet;
   b) holding said mirrored balls for concentrating solar energy somewhat rigidly captive in place in said sheet during and at the completion of its formation;
   c) introducing said infiltrating fluid to expand said sheet; and
   d) forming small fluid-filled individual annular cavities surrounding each said rotatable mirrored balls by the expansion caused by said infiltrating fluid.

2. The method according to claim **1**, wherein at least one dissolvable tray holds said mirrored balls in place in said sheet during its formation.

3. The method according to claim **1**, wherein at least one pillar supports said sheet to enhance fluid access during the sheet infiltration and expansion process.

4. The method according to claim **1**, wherein at least one of the fluids is optically transmissive.

5. The method according to claim **1**, wherein at least one of the fluids is a dielectric.

6. The method according to claim **1**, wherein the index of refraction of at least one of the fluids approximately matches that of said sheet.

7. The method according to claim **1**, wherein the density of at least one of the fluids approximately matches that of said mirrored balls.

8. The method according to claim **1**, wherein at least one of the fluids is lubricating.

9. The method according to claim **1**, wherein at least one of the infiltrating fluids is vaporously removed.

10. The method according to claim **1**, wherein the ratio of the overall volume of said sheet to the volume of said mirrored balls is between a factor 2 to 3.

11. The method according to claim **1**, wherein at least one monolayer of said rotatable mirrored balls is encapsulated in said sheet.

12. The method according to claim **1**, wherein more than one size of rotatable mirrored balls are encapsulated in said sheet.

13. The method according to claim **1**, wherein the mirrored balls are pre-coated, prior to being embedded in said sheet, to achieve minimal separation between the balls.

14. The method according to claim **1**, wherein the mirrored balls are asymmetrically closer to the top of said sheet than to the bottom.

15. The method according to claim **1**, wherein said sheet is constructed of bonded laminar films.

16. The method according to claim **1**, wherein zeolytes are in the fluid to help keep it clean and deionized.

17. A method of manufacturing a miniature optics, mirrored ball holding sheet comprising the fabrication stages of:
  a) dispersing a multitude of solid rotatable mirrored balls in a plasticizer mixture;
  b) forming a plastic sheet from said mixture which contains said balls with each ball individually encapsulated in the solidified sheet;
  c) introducing an infiltrating optically transmissive fluid that is absorbed more by said plastic sheet than by said balls; and
  d) emerging said sheet in said fluid causing the sheet to swell forming enlarged cavities that individually encapsulate each ball, wherein said miniature optics, mirrored ball holding sheet is a miniature optics, ball holding solar concentrator sheet.

18. The method of claim 17, wherein said fluid forms a lubricating shell surrounding each individually encapsulated ball.

19. The method of claim 17, wherein said fluid is a dielectric.

20. A method of fabricating a miniature optics, ball holding sheet comprising:
  a) dispersing a multitude of solid rotatable mirror balls in a hardenable fluid mixture;
  b) hardening said fluid mixture containing the mirrored balls; and
  c) providing an infiltrating liquid to expand the hardened mixture to form individual cavities surround each ball, wherein said miniature optics, ball holding sheet is a miniature optics, ball holding solar concentrator sheet.

21. The method of claim 20, wherein electromagnetic coupling means are provided to rotate said mirrored balls in an orientation responsive to an applied electromagnetic field.

22. The method of claim 20, wherein said liquid forms a lubricating shell surrounding each individually encapsulated ball.

* * * * *